\documentclass[12pt]{article}
\usepackage{bbm}
\usepackage{epsfig}
\usepackage{array}
\usepackage{float}
\usepackage{dsfont}
\usepackage{amstext}

\usepackage{a4}

\usepackage{a4wide}
\def\ra{\rightarrow}
\def\be{\begin{equation}}
\def\ee{\end{equation}}
\def\gs{\mathrel{
   \rlap{\raise 0.511ex \hbox{$>$}}{\lower 0.511ex \hbox{$\sim$}}}}
\def\ls{\mathrel{
   \rlap{\raise 0.511ex \hbox{$<$}}{\lower 0.511ex \hbox{$\sim$}}}}

\newcommand{\onbb}{neutrinoless double beta decay}
\newcommand{\ba}{\begin{array}{c}}
\newcommand{\baz}{\begin{array}{cc}}
\newcommand{\bad}{\begin{array}{ccc}}
\newcommand{\bea}{\begin{equation} \begin{array}{c}}
\newcommand{\eea}{ \end{array} \end{equation}}
\newcommand{\ea}{\end{array}}
\newcommand{\D}{\displaystyle}
\newcommand{\dms}{\mbox{$\Delta m^2_{\odot}$}}
\newcommand{\dma}{\mbox{$\Delta m^2_{\rm A}$}}
\newcommand{\meff}{\mbox{$\langle m \rangle$}}
\newcommand{\eV}{\mbox{ eV}}

\begin{document}

\title{\vspace{-2cm}
\vspace{-0.3cm} 
%\hfill {\small arXiv:0706.3YYY  [hep-ph]} 
%\vskip 1.72cm
\bf \Large
Inverted Mass Hierarchy from Scaling in the Neutrino Mass Matrix: 
Low and High Energy Phenomenology
}
\author{%\vskip -0.5cm
A.~Blum$^a$\thanks{email: \tt alexander.blum@mpi-hd.mpg.de}~\mbox{ 
},~R.~N.~Mohapatra$^b$\thanks{email: \tt rmohapat@physics.umd.edu}~\mbox{ 
},~W.~Rodejohann$^a$\thanks{email: \tt werner.rodejohann@mpi-hd.mpg.de}
\\\\
{\normalsize \it $^a$Max--Planck--Institut f\"ur Kernphysik,}\\
{\normalsize \it  Postfach 10 39 80, D--69029 Heidelberg, Germany}\\ \\ 
{\normalsize \it $^b$Department of Physics and Maryland Center for Fundamental Physics,}\\
{\normalsize \it University of Maryland, College Park, MD--20742, USA} \\
\small and \\
{\normalsize \it Sektion Physik, Ludwig--Maximilians--Universit\"at M\"unchen,} \\
{\normalsize \it Theresienstrasse 37a, D--80333  M\"unchen, Germany} \\
\small and \\
{\normalsize \it Physik--Department, Technische Universit\"at M\"unchen,}\\
{\normalsize \it  James--Franck--Strasse, D--85748 Garching, Germany} 
}
\date{}
\maketitle
\thispagestyle{empty}
\vspace{-0.8cm}
\begin{abstract}
\noindent
Best-fit values of recent global analyzes of neutrino data 
imply large solar neutrino mixing, vanishing $U_{e3}$ and a 
{\it non-maximal} atmospheric neutrino mixing angle $\theta_{23}$. We show that 
these values 
emerge naturally by the hypothesis of ``scaling'' in the Majorana 
neutrino mass matrix, which states that the ratios of its elements 
are equal. 
It also predicts an inverted hierarchy for the neutrino masses. 
We point out several advantages and distinguishing tests of the scaling 
hypothesis compared to the 
$L_e - L_\mu - L_\tau$ flavor symmetry, which is usually assumed 
to provide an 
understanding of the inverted hierarchy. 
Scenarios which have initially vanishing $U_{e3}$ and 
maximal atmospheric neutrino mixing are shown to be unlikely to 
lead to non-maximal $\theta_{23}$ while keeping simultaneously 
$U_{e3}$ zero.  
We find a peculiar ratio of the branching 
ratios $\mu \ra e \gamma$ and $\tau \ra e \gamma$ in supersymmetric 
seesaw frameworks, which only depends on atmospheric neutrino mixing 
and results in $\tau \ra e \gamma$ being unobservable.  
The consequences of the scaling hypothesis for high energy 
astrophysical neutrinos at neutrino telescopes are also investigated. 
Then we analyze a seesaw model based on the discrete symmetry 
$D_4 \times Z_2$ leading to scaling in the low energy mass matrix 
and being capable of generating the baryon asymmetry of 
the Universe via leptogenesis. The relevant CP phase is identical 
to the low energy Majorana 
phase and successful leptogenesis requires an effective mass 
for neutrinoless double beta decay larger than 0.045 eV.

\end{abstract}

\newpage

\section{\label{sec:intro}Introduction}
Observed lepton mixings are consequences of a non-trivial structure of 
the neutrino mass matrix ${\cal M}_\nu$. This symmetric matrix for 
Majorana neutrinos (having entries $m_{\alpha \beta}$ with 
$\alpha, \beta = e, \mu, \tau$) is in the charged lepton basis 
diagonalized by the Pontecorvo-Maki-Nakagawa-Sakata (PMNS) neutrino mixing 
matrix $U$. The very different structure of $U$  
compared to the quark sector for all possible neutrino mass orderings 
is indicative of an unexpected texture of the 
mass matrix, and could hold important clues to our understanding of the 
physics of fundamental constituents of matter. To unravel this new 
physics, various Ans\"atze 
for ${\cal M}_\nu$ have been made in the literature \cite{reviews} and 
their associated symmetries have been sought after. 
One particular proposal, recently proposed by two of us (R.N.M.~and W.R.), 
on which we will focus in this note, is called 
``scaling'' \cite{MR}.  
The scaling hypothesis demands that the ratio 
$\frac{m_{\alpha \beta}}{m_{\alpha \gamma}}$ is  
independent of the flavor $\alpha$: 
\be \label{eq:0}
\frac{m_{e \beta}}{m_{e \gamma}} = 
\frac{m_{\mu \beta}}{m_{\mu \gamma}} = 
\frac{m_{\tau \beta}}{m_{\tau \gamma}} = 
c ~\mbox{ for fixed $\beta$ and $\gamma$}~.	 
\ee
There are three possibilities and the only one phenomenologically allowed 
is when $\beta = \mu$ and $\gamma = \tau$. We shall call this 
case scaling henceforth. The resulting mass matrix reads   
\be \label{eq:mnustrong}
{\cal M}_\nu = m_0 \, 
\left( 
\bad
A & B & B/c \\[0.2cm]
B & D & D/c \\[0.2cm]
B/c & D/c & D/c^2 
\ea
\right)~.
\ee
Similar matrices have been found independently in the context of specific 
models (see Ref.~\cite{others}). The most important phenomenological 
prediction of scaling is that Eq.~(\ref{eq:mnustrong}) leads 
to an inverted hierarchy with $m_3 = 0$ and $U_{e3}=0$. 
Atmospheric neutrino mixing is 
governed by the ``scaling factor'' $c$ via 
$\tan^2 \theta_{23} = 1/c^2$, i.e., is in general non-maximal 
because $c$ is naturally of order, but not equal to, one. 
It is interesting to note that current data analyzes 
(though at the present stage statistically not very significant) 
yield non-maximal 
$\tan^2 \theta_{23} = 0.89$ as the best-fit point \cite{data} 
(see also \cite{lisi}, where the best-fit value 
is $\tan^2 \theta_{23} = 0.82$). The reason is that in the SuperKamiokande 
experiment there is an excess of sub-GeV electron events, but 
no excess either of sub-GeV muon events or of multi-GeV 
electrons. In a realistic 3-flavor analysis this prefers 
$\cos \theta_{23} > \sin \theta_{23}$ \cite{data,lisi}. 
The best-fit value of $U_{e3}$ is zero.

If these 
values, $\theta_{23} \neq \pi/4$ and $U_{e3}=0$, 
are indeed confirmed by future data 
then one should look for symmetries and/or models which are capable 
of predicting such a situation. Ideally, such a candidate should 
be rather insensitive to radiative corrections and should 
not require much, if any, breaking to achieve the values sought for. 
Scaling is one such appealing possibility, several general 
aspects of which will be discussed in Section \ref{sec:gen}. 
Typical models are shown to predict -- when constructed in 
a supersymmetric seesaw framework -- a characteristic 
ratio of the branching 
ratios $\mu \ra e \gamma$ and $\tau \ra e \gamma$. 
The latter decay is then too rare to be observable in presently 
foreseen experiments. Simple phenomenology of fluxes 
of high energy astrophysical neutrinos at neutrino telescopes 
is predicted and studied in Section \ref{sec:UHE}. 
We argue further in Section \ref{sec:comp0} 
that it is difficult 
to obtain $\theta_{23} \neq \pi/4$ and $U_{e3}=0$ in scenarios 
in which initially $\theta_{23} = \pi/4$ and $U_{e3}=0$ holds. 
Stressing that 
scaling predicts an inverted hierarchy leads us to compare  
the Ansatz with the flavor symmetry $L_e - L_\mu - L_\tau$ 
\cite{lelmlt}. 
The latter is usually assumed to be the origin of an inverted hierarchy. 
We show in Section \ref{sec:comp} that scaling possesses 
several advantages over $L_e - L_\mu - L_\tau$. 
If future 
experiments indeed show that neutrinos obey an inverted hierarchy, 
then one needs a full list of possible scenarios that can predict it. 
Necessarily, these are unusual symmetries as typical GUT models do 
not lead to an inverted ordering. Accordingly, we 
investigate in Section \ref{sec:mod} 
a seesaw model leading to scaling based on the discrete symmetry 
$D_4\times Z_2$. 
We show that the baryon asymmetry of the Universe 
via leptogenesis can be reproduced 
and analyze the connection to the low 
energy parameters. We summarize in Section \ref{sec:concl}.

\section{\label{sec:gen}General Properties of Scaling}

The neutrino mass matrix giving rise to scaling is given in 
Eq.~(\ref{eq:mnustrong}). It appears in the Lagrangian 
\be \label{eq:L}
{\cal L} = \frac 12 \, \overline{\nu_L^c} \, {\cal M}_\nu \, \nu_L + 
\overline{\ell_R} \, {\cal M}_\ell \, \ell_L ~, 
\ee
where ${\cal M}_\ell$ is the charged lepton mass matrix. 
Diagonalizing it via 
${\cal M}_\ell = V_\ell \, {\cal M}_\ell^{\rm diag} \, 
U_\ell^\dagger$ and the neutrino mass matrix with 
$U_\nu^\ast \, {\cal M}_\nu^{\rm diag} \, U_\nu^\dagger = {\cal M}_\nu$, 
gives us the PMNS matrix 
\bea \label{eq:Upara}
U = U_\ell^\dagger \, U_\nu = 
\left( \bad 
c_{12} \, c_{13} & s_{12}  \, c_{13} & s_{13} \, e^{-i \delta} \\[0.2cm] 
-s_{12}  \, c_{23} - c_{12}  \, s_{23}  \, s_{13} \, e^{i \delta}  
& c_{12}  \, c_{23} - s_{12}  \, s_{23}  \, s_{13} \, e^{i \delta}
& s_{23}  \, c_{13}  \\[0.2cm] 
s_{12}  \, s_{23} - c_{12}  \, c_{23}  \, s_{13}  \, e^{i \delta}& 
- c_{12}  \, s_{23} - s_{12}  \, c_{23}  \, s_{13}  \, e^{i \delta}
& c_{23}  \, c_{13}  \\ 
               \ea   \right) \, P ~, 
\eea
where $c_{ij} = \cos\theta_{ij}$, 
$s_{ij} = \sin\theta_{ij}$, and 
$P = {\rm diag}(1,e^{i \alpha},e^{i (\beta + \delta)})$ contains the 
Majorana phases \cite{MajPha}. The best-fit values as well as the 
allowed $1\sigma$ and $3\sigma$ ranges of the 
oscillation parameters are \cite{data}: 
\begin{eqnarray} \label{eq:data}
\dms &=& 
\left(7.9^{+0.3\,, \,1.0}_{-0.3\,, \,0.8}\right) 
\cdot 10^{-5} \eV^2~,\nonumber\\
\sin^2 \theta_{12}
 &=& 0.31^{+0.02\,, \,0.07}_{-0.02\,, \,0.06} ~,\nonumber\\
\dma &=&  
\left(2.6^{+0.2\,, \,0.6}_{-0.2\,, \,0.6}\right) 
\cdot 10^{-3} \eV^2~,\\
\tan^2\theta_{23} &=& 0.89^{+0.31\,, \,0.89}_{-0.21\,, \,0.42} ~,\nonumber\\
|U_{e3}|^2 &<& 0.008~(0.040)~,\nonumber
\end{eqnarray}
where $\dms = m_2^2 - m_1^2$ and  $\dma = |m_3^2 - m_1^2|$.\\

We will most of the 
time assume that scaling holds in the 
charged lepton basis, i.e., $U_\ell = \mathbbm{1}$. 
We nevertheless stress here the following interesting 
property of scaling: if $U_\ell$ is non-trivial, then 
in the charged lepton basis we have 
$\tilde {\cal M}_\nu = U_\ell^T \, {\cal M}_\nu \, U_\ell$. In case 
$U_\ell$ is only given by a 23-rotation it is easy to 
show that the $ee$ entry of ${\cal M}_\nu$ is not affected, 
and in addition we have:
\be \label{eq:c23}
\frac{\tilde m_{e \mu}}{\tilde m_{e \tau}} = 
\frac{\tilde m_{\mu \mu}}{\tilde m_{\mu \tau}} = 
\frac{\tilde m_{\mu \tau}}{\tilde m_{\tau \tau}} = 
\frac{ c \, \cos \theta_{23}^\ell - \sin \theta_{23}^\ell}
{\cos \theta_{23}^\ell + c \, \sin \theta_{23}^\ell} \equiv \tilde c~,
\ee
where $\theta_{23}^\ell$ is the rotation angle of $U_\ell$ and 
$\tilde m_{\alpha \beta}$ are the entries of $\tilde {\cal M}_\nu$. 
Consequently, the texture of the neutrino mass matrix is not 
changed at all, it still obeys scaling and predicts $m_3 = U_{e3} = 0$ 
but now with $\tan^2 \theta_{23} = 1/\tilde{c}^2$.  
If for some reason in ${\cal M}_\nu$ the scaling factor $c$ is 
much larger or smaller than 1, then a 23-rotation from the 
charged lepton sector can still save the scaling hypothesis. 
For $c \gg 1$ we find $\tan^2 \theta_{23} \simeq \tan^2 \theta_{23}^\ell$, 
while for $c \ll 1$ we have 
$\tan^2 \theta_{23} \simeq \cot^2 \theta_{23}^\ell$. 
This observation can simplify the construction of models.\\

Interestingly, the characteristic predictions $m_3 = U_{e3} = 0$
are not subject to any radiative corrections when going 
from high scale down to low scale. This can be understood by 
letting RG effects directly modify the mass matrix, 
as done in Ref.~\cite{MR}, 
or by glancing at the RG equations of $\theta_{13}$ and $m_3$. 
For both of them it holds that \cite{RGEs} 
\be
\dot m_3 \propto m_3~~\mbox{ and }~~ \dot \theta_{13} \propto m_3~,
\ee
i.e., $\theta_{13} = m_3 = 0$ is stable under RG evolution.

What are other phenomenological predictions of scaling? 
First of all, there 
will be no CP violation in oscillation experiments because 
of $\theta_{13}=0$. Then we note that from $m_3=0$ it follows that 
one Majorana phase is unphysical. The other one appears in  
the effective mass to which \onbb~is sensitive \cite{0vbb_APS}. 
This parameter 
takes a very simple form for the inverted hierarchy with 
$m_3 = \theta_{13} = 0$:
\be
\meff = \sqrt{\dma} \, \sqrt{1 - \sin^2 2 \theta_{12} \, \sin^2 \alpha}~.
\ee 
The range of \meff~lies for the best-fit parameters from 
Eq.~(\ref{eq:data}) between 0.019 and 0.051 eV, 
while at 1(3)$\sigma$ it ranges between 0.017 and 0.053 eV 
(0.011 and 0.057 eV). If the parameters do not conspire to render 
\meff~at the very low end of this range then next-generation 
experiments will definitely observe \onbb~\cite{0vbb_APS}. The conditions 
under which one can extract $\alpha$ from an observation 
of \onbb~are given in Ref.~\cite{CP_0vbb}. Unlike many 
other approaches, the scaling Ansatz can therefore be completely 
reconstructed. Of course, the scaling Ansatz is  
easier to disprove than to prove. However, if 
future experiments give very strong limits on $|U_{e3}|$ and the inverted
hierarchy is present, then this would be a very strong hint towards the
realization of scaling.\\

How can a low energy mass matrix like Eq.~(\ref{eq:mnustrong}) 
be achieved? We work of course in the framework of the seesaw mechanism 
\cite{seesaw} in which ${\cal M}_\nu = -M_D^T \, M_R^{-1} \, M_D$, where 
$M_D$ is the Dirac and $M_R$ the heavy Majorana neutrino mass matrix. 
One remarkable property of scaling is the 
following: if the Dirac mass matrix obeys 
scaling, i.e., 
\be \label{eq:md0}
M_D = 
\left( 
\bad 
a_1 & b & b/c \\
a_2 & d & d/c \\
a_3 & e & e/c 
\ea
\right) ~,
\ee
then ${\cal M}_\nu$ takes the form obeying scaling\footnote{Note 
that there is no need for $M_R$ to obey scaling because it would 
be singular in this case, invalidating simple seesaw.}
 from Eq.~(\ref{eq:mnustrong}) 
{\it regardless of the structure of $M_R$}! Note that $M_D$ is not necessarily 
required to be symmetric.

Within supersymmetrized seesaw models one has an interesting 
connection to lepton flavor violating (LFV) decays of charged leptons 
such as $\mu \ra e \gamma$ \cite{BRs}. RG evolution within theories of 
universal boundary (mSUGRA) conditions leads to off-diagonal 
entries in the slepton mass matrix, which trigger effects of LFV. 
In particular, branching ratios of the decays 
$\ell_i \ra \ell_j \gamma$ with $(\ell_i, \ell_j) =
 (\mu, e),~(\tau, e),~(\tau, \mu)$ are 
given by \cite{BRs}
\bea \D \label{eq:BRs}
\frac{ {\rm BR}(\ell_i \ra \ell_j \gamma)}
{{\rm BR}(\ell_i \ra e \, \nu \, \overline{\nu})} 
=  \frac{\alpha^3}{G_F^2 \, v_{\rm wk}^4 \, m_S^8} \, \left| 
\frac{(3 + a_0^2) \, m_0^2}{8 \, \pi^2} \right|^2 \, 
\left| (M_D^\dagger \, L \, M_D)_{ij} \right|^2 \, \tan^2 \beta~, \\[0.2cm]
\D \mbox{ where } 
(L)_{ij} = \delta_{ij} \ln \frac{M_X}{M_i}~.
\eea
Here $v_{\rm wk} = 174$ GeV, $M_i$ are the heavy Majorana neutrino 
masses and 
$M_X > M_i$ is the scale at which universal boundary conditions 
are implemented. The SUSY 
parameters are $a_0 = A_0/m_0$ with $m_0$ the universal scalar mass,  
$A_0$ the universal trilinear coupling, and $m_S$ is a typical SUSY mass. 
Neglecting the only logarithmic dependence 
on the heavy Majorana masses, the branching ratios are   
proportional to the modulus-squared of the off-diagonal entries 
of $M_D^\dagger \, M_D$. It follows with Eq.~(\ref{eq:md0}) that  
\be \label{eq:LFV_rel}
\frac{1}{{\rm BR}(\tau \ra e \, \nu \, \overline{\nu})}
\frac{{\rm BR}(\tau \ra e \gamma)}{{\rm BR}(\mu \ra e \gamma)} 
= 
\left| \frac{(M_D^\dagger \, M_D)_{31}}
{(M_D^\dagger \, M_D)_{21}} \right|^2 
= \frac{1}{c^2} = \tan^2 \theta_{23}~, 
\ee 
with ${\rm BR}(\tau \ra e \, \nu \, \overline{\nu}) \simeq 0.1784$. 
The two branching ratios are therefore simply related by the atmospheric 
neutrino mixing angle. 
Consequently, such models predict them within (see Eq.~(\ref{eq:data})) 
a factor of less than four. The current limit of  
${\rm BR}(\mu \ra e \gamma) \le 1.2 \times 10^{-11}$ \cite{meg_lim}, 
implies therefore that 
${\rm BR}(\tau \ra e \gamma)$ will always be close to this number and 
consequently at least two orders of magnitude below the future limits 
(between $10^{-8}$ and $10^{-9}$) which are currently foreseen.  

To be more precise, the branching ratios as defined in 
Eq.~(\ref{eq:BRs}) have to be evaluated in 
the basis in which the charged leptons and the heavy 
Majorana neutrinos are diagonal. In this case $M_D$ has to be 
replaced with $\tilde{M}_D = V_R^T \, M_D \, U_\ell$, 
where $M_R = V_R^\ast \, M_R^{\rm diag} \, V_R^\dagger$. 
Obviously, $V_R$ drops out of $\tilde{M}_D^\dagger \, \tilde{M}_D$ 
and does not influence LFV.  
Now consider again the case that $U_\ell$ is non-trivial and 
given by a 23-rotation, which we showed above to keep the 
scaling predictions of $m_3 = U_{e3} = 0$ and to change 
$c$ to $\tilde{c}$ given in Eq.~(\ref{eq:c23}). 
One easily finds that
\be
\left| \frac{(\tilde{M}_D^\dagger \, \tilde{M}_D)_{31}}
{(\tilde{M}_D^\dagger \, \tilde{M}_D)_{21}} \right|^2 = 
\left|\frac{\cos \theta_{23}^\ell + c \, \sin \theta_{23}^\ell}
{ c \, \cos \theta_{23}^\ell - \sin \theta_{23}^\ell}
 \right|^2~.
\ee
This expression is nothing but $1/\tilde{c}^2$ and therefore we 
recover the relation Eq.~(\ref{eq:LFV_rel}) between the ``double ratio'' 
${\rm BR}(\tau \ra e \gamma)/{\rm BR}(\mu \ra e \gamma)$ and atmospheric 
neutrino mixing.\\ 

It is interesting to ask whether scaling can be applied to non-standard 
neutrino scenarios. In particular, the possibility of neutrinos 
being Dirac neutrinos and the presence of additional light sterile 
neutrino species is discussed frequently. 

Dirac neutrinos could be accommodated by scaling if the 
neutrino mass matrix would take the form given in Eq.~(\ref{eq:md0}). 
The resulting neutrino oscillation 
phenomenology (obtained by diagonalizing $M_D^\dagger \, M_D$)  
would be again an inverted hierarchy with 
$\tan^2 \theta_{23} = 1/c^2$ and $U_{e3}= m_3 = 0$. The reason is simply 
because the resulting mass matrix possesses an eigenvalue 
0 with a corresponding eigenvector of $(0,-1/c,1)^T$. 

Sterile neutrinos would require enlarging ${\cal M}_\nu$ from 
being a $3\times3$ matrix to a $(3 + n_s)\times(3+n_s)$ matrix, 
where $n_s$ is the number 
of additional sterile neutrinos. The recent results of the MiniBooNE 
experiment \cite{mini} seem to indicate that $n_s \ge 2$ \cite{3+2}. 
We can modify the scaling condition from Eq.~(\ref{eq:0}) to 
include also the sterile neutrinos: 
\be
\frac{m_{e\mu}}{m_{e\tau}} = \frac{m_{\mu\mu}}{m_{\mu\tau}} =
\frac{m_{\tau \mu}}{m_{\tau\tau}} = 
\frac{m_{ s_1 \mu}}{m_{s_1 \tau}} = \frac{m_{ s_2 \mu}}{m_{s_2 \tau}}
=\ldots = c~.
\ee
The result is a mass matrix with a zero eigenvalue having an  
eigenvector $(0,-1/c,1,0,0,\ldots)^T$. 
The mixing scenario is described by $U_{e3}=0$, 
$|U_{\mu 3}/U_{\tau 3}|^2 = 1/c^2$ and $U_{s_1 3} = U_{s_2 3} 
= \ldots = 0$. Leptonic CP violation is important in order to 
allow scenarios with two sterile neutrinos to survive constraints 
from current data \cite{3+2}. In total there are five CP phases 
in this case, and only three of them are unphysical due to the 
vanishing mixing matrix elements. 
For two sterile neutrinos, the resulting scenario would correspond 
to an inverted hierarchy of 
the three mostly active neutrinos and two heavier, mostly sterile 
neutrinos. It is in the language of Ref.~\cite{GR} the scenario ``SSI'', 
which has from all possible mass orderings the smallest predictions 
for the various mass-related observables (neutrinoless double 
beta decay, neutrino mass in KATRIN and the sum of masses in cosmology) 
\cite{GR}.

%%%%%%%%%%%%%%%%%%%%%%%%%%%%%%%%%%%%%%%%%%%%%%%%%%%%%%%%%%%%%%%%%%%%%%%%%
\section{\label{sec:UHE}Scaling Predictions for Astrophysical Neutrinos}
%%%%%%%%%%%%%%%%%%%%%%%%%%%%%%%%%%%%%%%%%%%%%%%%%%%%%%%%%%%%%%%%%%%%%%%%

It has recently been recognized that measuring flux ratios of high energy 
astrophysical neutrinos \cite{UHEs} is an alternative method 
to determine neutrino mixing parameters \cite{UHE,xing,WR}. In particular, 
one expects from astrophysical $pp$ or $p\gamma$ 
processes, which generate pions and kaons, an initial 
flux composition of the form $\Phi_e^0 : \Phi_\mu^0 : \Phi_\tau^0 = 
1 : 2 : 0$, 
where $\Phi_\alpha^0$ with $\alpha = e, \mu, \tau$ is the 
flux of neutrinos 
with flavor $\alpha$. Neutrino mixing modifies the flavor 
composition and in terrestrial neutrino telescopes such as 
IceCube \cite{ice} one can then measure flux ratios and thereby 
obtain information on the neutrino parameters. The measurable 
neutrino flux is given by 
\be
\Phi_\alpha = \sum\limits_\beta P_{\alpha \beta}\, \Phi_\alpha^0
~\mbox{ with } P_{\alpha \beta} = \sum\limits_i |U_{\alpha i}|^2 \, 
|U_{\beta i}|^2~.
\ee
In the limit of maximal atmospheric neutrino 
mixing and vanishing $U_{e3}$, the composition $1 : 2 : 0$ 
is transformed into 
$1 : 1 : 1$, independent of the solar neutrino mixing angle. Small 
deviations from $\theta_{23} = \pi/4$ and $\theta_{13} = 0$ lead to 
\cite{xing,WR}
\bea \label{eq:Delta} 
\Phi_e : \Phi_\mu : \Phi_\tau = 1 + 2 \, \Delta : 1 - \Delta : 1 - \Delta ~,
\\[0.2cm] \D 
\mbox{ where } \Delta = \frac 14 \, \sin 4 \theta_{12} \, 
|U_{e3}| \, \cos \delta + \frac 12 \, \sin^2 2 \theta_{12} 
\, \left(\frac 12 - \sin^2 \theta_{23} \right)~.
\eea
Thus, there is a universal first order correction in terms of 
the small parameters $|U_{e3}|$ and $\frac 12 - \sin^2 \theta_{23}$. 
With the current $1\sigma$ ($3\sigma$) ranges of the 
oscillation parameters one finds that 
$-0.036 \le \Delta \le 0.057$ $(-0.097 \le \Delta \le 0.113)$. 
The ratio of electron 
neutrinos to the other flavors is therefore a probe of $\theta_{23}$, 
$\theta_{13}$ and $\cos \delta$.  
Note that in the definition of $\Delta$ in Eq.~(\ref{eq:Delta}) 
the factor in front of $\frac 12 - \sin^2 \theta_{23}$ is larger 
and has a smaller range 
than the one in front of $|U_{e3}| \, \cos \delta$. 
%Moreover, the 
%range of the respective factor is considerably smaller for 
%$\frac 12 - \sin^2 \theta_{23}$. 
To be precise, for the 
allowed $3\sigma$ range of solar neutrino mixing, 
$\frac 14  \sin 4 \theta_{12}$ ranges from 0.12 to 0.21, whereas 
$\frac 12  \sin^2 2 \theta_{12} $ ranges from 0.38 to 0.48. Consequently 
\cite{WR}, the sensitivity to deviations from maximal 
atmospheric neutrino mixing 
is better than the sensitivity to deviations from $|U_{e3}|=0$, 
which in addition gets smeared by 
the dependence on the CP phase $\delta$. 

\begin{figure}[ht]
\centerline{\psfig{figure=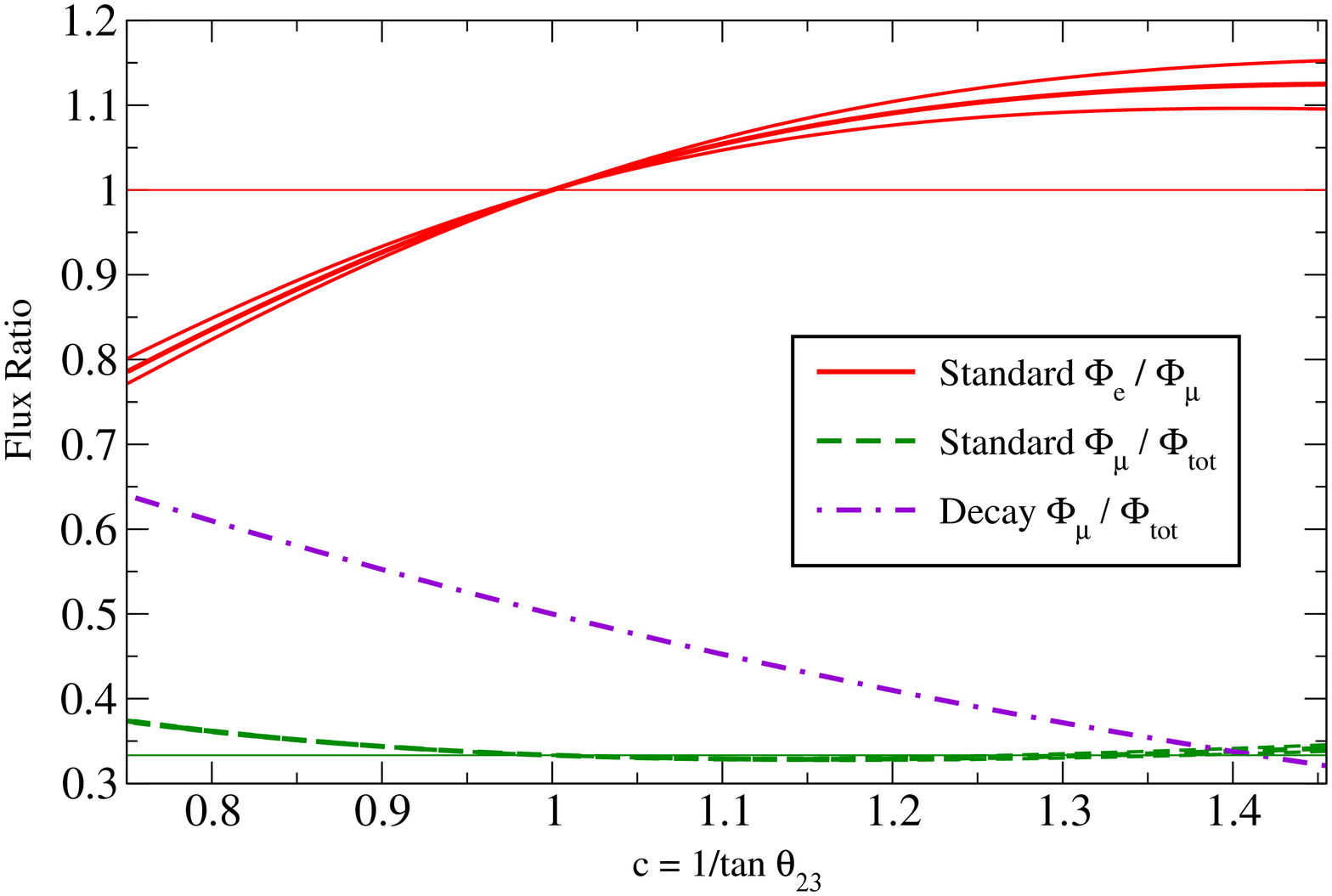,height=8cm,width=12cm}}
\caption{\label{fig:UHE}Flux ratios of high energy astrophysical 
neutrinos as a function of the scaling parameter $c$. The red (solid) 
lines are $\Phi_e/\Phi_\mu$ for the standard case of an initial 
$1 : 2 : 0$ composition, the green (dashed) lines 
are $\Phi_\mu/\Phi_{\rm tot}$ in the standard case and the violet 
(dot-dashed) line is for neutrino decay. The first two cases are 
plotted for the best-fit values and the upper and lower $3\sigma$ values 
of the solar neutrino mixing angle $\theta_{12}$.}
\end{figure}

We conclude that scaling -- predicting only non-maximal $\theta_{23}$ 
and neither $\delta$ nor $\theta_{13}$ -- will have particularly simple, 
interesting and potentially testable phenomenology at neutrino telescopes. 
To go into more detail, let us introduce the small parameter 
\be \nonumber 
\epsilon = \frac{\pi}{4} - \theta_{23} ~,
\ee
which can be linked to the scaling parameter $c$ via 
$\frac 12 (c^2 - 1)/(c^2 + 1) = \frac 12 - \sin^2 \theta_{23} 
\simeq \epsilon $.   
The result for the flux ratios is 
\bea \nonumber 
\Phi_e : \Phi_\mu : \Phi_\tau = \\[0.2cm] 
\hspace{-.2cm}
\mbox{\small $1 + 2 \, c_{12}^2 \, s_{12}^2 \, (c_{23}^2 - s_{23}^2) : 
2  \left (1 - 2 \, c_{23}^2 \, s_{23}^2  - c_{12}^2 \, s_{12}^2 \, 
(c_{23}^2 - s_{23}^2 ) \, c_{23}^2  \right) : 
2 \, s_{23}^2 \left( 1 + (1 - c_{12}^2 \, s_{12}^2) 
(c_{23}^2 - s_{23}^2) \right)  $}\\[0.2cm] 
\simeq 
1 + 4 \, c_{12}^2 \, s_{12}^2 \, \epsilon : 
1 - 2 \, c_{12}^2 \, s_{12}^2 \, \epsilon 
: 1 - 2 \, c_{12}^2 \, s_{12}^2 \, \epsilon ~,
\eea 
where we have given the exact expression and the expansion in 
terms of $\epsilon$ to first order. The electron neutrino flux 
$\Phi_e$ receives no quadratic correction and the second order term 
for $\Phi_\mu $ is $4 \, (1 - c_{12}^2 \, s_{12}^2) \, \epsilon^2$, 
while that 
for $\Phi_\tau$ is identical but with opposite sign. 
In Fig.~\ref{fig:UHE} we plot -- using the exact probabilities -- 
the flux ratios $\Phi_e/\Phi_\mu$ and 
$\Phi_\mu/\Phi_{\rm tot}$, where $\Phi_{\rm tot}$ is the total 
neutrino flux, as a function of the scaling parameter $c$. 
For $c = 1$ we obtain  
$\Phi_e/\Phi_\mu = 1$ and $\Phi_\mu/\Phi_{\rm tot} = \frac 13$. 
Quite large deviations from these values are allowed, and the 
dependence on solar neutrino mixing is very weak. 

Finally, we note that if there are sufficiently fast non-standard decay modes 
of the neutrinos, and only the lightest state $\nu_i$ ($i = 1$ for normal 
ordering, $i = 3$ for inverted ordering) survives and is detected, 
the fluxes obey the relation \cite{decay} 
$\Phi_e : \Phi_\mu : \Phi_\tau = 
|U_{e i}|^2 : |U_{\mu i}|^2 : |U_{\tau i}|^2$. In case 
of scaling (inverted hierarchy and $\theta_{13}=0$), this simplifies 
to   
\be
\Phi_e : \Phi_\mu : \Phi_\tau = 
0 : \sin^2 \theta_{23} : \cos^2 \theta_{23} = 0 : 1 : c^2~.
\ee 
There are no electron 
neutrinos and the ratio of muon to tau neutrinos is 
$\tan^2 \theta_{23} = 1/c^2$. We plot for the case of decaying 
neutrinos in Fig.~\ref{fig:UHE} the ratio 
of muon neutrinos to the total flux, which is simply equal 
to $\sin^2 \theta_{23}$.

\section{\label{sec:comp0}Non-maximal atmospheric Neutrino Mixing 
and vanishing $U_{e3}$ from other Scenarios}

The question arises if we can obtain the values $\theta_{13}=0$ 
and $\theta_{23} \neq \pi/4$ by breaking or modifying scenarios 
with initial $\theta_{13}=0$ and $\theta_{23} = \pi/4$. 
Explicit breaking of the symmetry in a mass matrix, 
RG effects and contributions from the 
charged lepton sector are appealing possibilities, which we will now 
comment on. 

Considering first radiative corrections, 
we note here that $\dot \theta_{13}$ and $\dot \theta_{23}$ 
are inversely proportional to \dma~\cite{RGEs}, 
and due to this one would expect 
that in general RG corrections to them are of the 
same order and in addition small. Therefore, if initially $\theta_{13} = 
\theta_{23} - \pi/4 = 0$, then keeping at low energy 
$\theta_{13}$ zero but having simultaneously 
$\theta_{23}$ non-maximal would require 
rather special values of the other parameters 
and unnatural cancellations in particular between the CP phases.  
Hence, $\theta_{23} \neq \pi/4$ and $\theta_{13}=0$ should hold 
initially and we end up again with the 
question of how these peculiar values arose, which leads us 
back to the scaling hypothesis.

Turning to explicit breaking of the symmetry in a mass matrix leading 
to $\theta_{13} = 0$ and $\theta_{23} = \pi/4$ 
requires a glance at $\mu$--$\tau$ symmetry \cite{mutau}, 
which imprints the following form on the mass matrix: 
\be \label{eq:mutau}
{\cal M}_\nu = m_0 \, 
\left( 
\bad 
A & B & B \\[0.2cm] 
B & D & E \\[0.2cm] 
B & E & D 
\ea 
\right)~. 
\ee
Maximal $\theta_{23}$ and zero $U_{e3}$ are predicted, but only if 
the eigenvalue belonging to the eigenvector 
$(0,-1,1)^T$ is the largest or smallest one, which would then correspond to 
the normal or inverted mass ordering.  
In case of scaling, the eigenvector $(0,-1/c,1)^T$ belongs 
automatically to the zero eigenvalue, therefore there is 
no such ambiguity. 
For a normal hierarchy of the neutrino masses 
the parameters in Eq.~(\ref{eq:mutau}) 
have to fulfill $D,E \gg A,B$. If in this case 
the $\mu$--$\tau$ symmetry is broken such that only the $\mu\mu$ and 
$\mu\tau$ entries differ, but the $e\mu$ and $e\tau$ elements stay 
identical, then it turns out that 
$U_{e3}$ is small but non-zero (to be precise, it is of order 
$\dms/\dma$), while $\theta_{23}$ deviates from maximal by roughly 
$\sqrt{\dms/\dma}$ \cite{rabimutau}. This way of breaking has 
no analogue for an inverted hierarchy 
or quasi-degenerate neutrinos, for which rather tuned breaking scenarios 
are required in order to end up with 
$\theta_{13} = 0$ but $\theta_{23} \neq \pi/4$.

Contributions from the charged leptons arise when  
${\cal M}_\nu$ is $\mu$--$\tau$ symmetric and in the 
limit of a diagonal charged lepton mass matrix 
$\theta_{13} = 0$ and $\theta_{23} = \pi/4$ would result from 
$U = U_\ell^\dagger \, U_\nu = U_\nu$. 
It is commonly assumed that $U_\ell$ contains only small angles. 
Introducing for the sines of these angles the abbreviations 
$\sin \theta_{ij}^\ell \equiv \lambda_{ij}$, one finds 
in first order of these small parameters that \cite{Ulep,FPR} 
\bea \label{eq:HPR}
|U_{e3}| \simeq \frac{1}{\sqrt{2}} \, 
\left| 
\lambda_{12} - \lambda_{13} \, e^{i \phi_1} 
\right|
~,\\[0.2cm]
\sin^2 \theta_{23} \simeq \frac 12 + \lambda_{23} 
\, \cos \phi_2 - 
\frac 14 \left(\lambda_{12}^2  - \lambda_{13}^2 \right) 
+ \frac 12 \, \cos \phi_1 \, 
\lambda_{12} \, \lambda_{13}~,
\eea
where $\phi_1$ and $\phi_2$ are CP phases. 
Consequently, $\theta_{13} = 0$ and $\theta_{23} \neq \pi/4$ would 
require delicate interplay of the angles and phases in $U_\ell$ and would 
lead to a rather unnatural form for it. 
In particular, for the natural case of $U_\ell$ being CKM-like, i.e., 
$\lambda_{12} \gg \lambda_{13,23}$, the result 
$\theta_{13} = 0$ and $\theta_{23} \neq \pi/4$ can 
not be achieved.  Similar statements hold for 
the opposite case in which in the limit $U_\nu = \mathbbm{1}$ 
the charged lepton sector would suffice to generate 
$\theta_{13} = 0$ and $\theta_{23} = \pi/4$ in 
$U = U_\ell^\dagger \, U_\nu = U_\ell^\dagger$ \cite{Ulep}.\\

We conclude that it seems rather unnatural to obtain $U_{e3}=0$ and 
$\theta_{23} \neq \pi/4$ within broken $\mu$--$\tau$ symmetry. 
These predictions for the mixing angles also occur in models based 
on the flavor symmetry $L_e - L_\mu - L_\tau$. However, as we will show 
in the next Section, various tuning problems show up for this Ansatz.

\section{\label{sec:comp}Inverted Hierarchy: 
Scaling vs.~the $L_e - L_\mu - L_\tau$ Flavor Symmetry}

Note that scaling is an Ansatz for 
the inverted hierarchy which 
is fundamentally different from the flavor symmetry 
$L_e - L_\mu - L_\tau$ \cite{lelmlt}, which is usually ``blamed'' 
for it. A detailed comparison is therefore a worthy exercise. 
A mass matrix obeying the flavor symmetry $L_e - L_\mu - L_\tau$ reads 
\be \label{eq:lelmlt}
{\cal M}_\nu = m_0 \, 
\left( 
\bad
0 & \cos \theta_{23} & -\sin \theta_{23} \\[0.2cm]
\cos \theta  & 0 & 0 \\[0.2cm]
-\sin \theta_{23} & 0 & 0 
\ea
\right)\,,\mbox{ where } m_0 = \sqrt{\dma}~,
\ee
and generates $m_3 = U_{e3} = 0$ as well as 
non-maximal atmospheric neutrino mixing given by 
$\theta_{23}$. However, it also predicts 
{\it maximal} solar neutrino mixing and vanishing $\dms = m_2^2 - m_1^2$, 
which is in contradiction to observation.  
Therefore, in contrast to scaling, the symmetry needs to be broken 
to achieve correct phenomenology, 
which imposes three problems \cite{lelmlt1}: 
\begin{itemize}
\item [(i)] the breaking terms in the mass matrix have to have at least 
$30 \%$ the magnitude of the terms allowed by the symmetry. The reason 
is that in the inverted hierarchy the $ee$ entry of ${\cal M}_\nu$ 
(the effective mass) is required to be larger 
than\footnote{The fact that 
$L_e - L_\mu - L_\tau$ predicts an effective mass near the lower end 
of its allowed range, whereas scaling admits \meff~to take any of its 
allowed values, represents a possibility to distinguish the two Ans\"atze.}
$\sqrt{\dma} \, \cos 2 \theta_{12}$; 

\item [(ii)] the (large) breaking of the symmetry in the mass 
matrix is always connected with fine-tuning because 
usually the required large deviation from maximal 
$\theta_{12}$ is connected with the small ratio of the solar
and atmospheric mass-squared differences. For instance, let us 
add to the matrix in Eq.~(\ref{eq:lelmlt}) the 
following ($\mu$--$\tau$ symmetric) perturbation:  
\[ 
\epsilon ~
\left( 
\bad
a & b & b \\ 
b & d & e \\
b & e & d 
\ea
\right)~.
\]   
If for simplicity we set $\theta_{23} = \pi/4$ in Eq.~(\ref{eq:lelmlt}) 
then maximal atmospheric neutrino mixing and $U_{e3}=0$ 
will remain, but the ratio of mass-squared 
differences is now $\dms/\dma \simeq \sqrt{2} \, (a + d + e) \, \epsilon$ 
while solar neutrino mixing is governed by 
$\sin \theta_{12} \simeq 1/\sqrt{2} - (a - d - e)\,\epsilon/8$. 
Hence, if one wants to reproduce the best-fit values of the oscillation 
parameters in Eq.~(\ref{eq:data}), the fine-tuned condition 
$(a + d + e)/(a - d - e) \simeq 0.0027$ has to be fulfilled; 

\item [(iii)] if the symmetry is 
broken by contributions from the charged lepton 
sector (note that in this case in addition breaking 
in the neutrino sector is necessary 
to generate non-vanishing \dms) 
a CP violating phase appears in the expression for the 
now non-maximal solar neutrino mixing angle, which is 
required to be close to zero: 
using again the natural choice of a CKM-like $U_\ell$ 
(see also Eq.~(\ref{eq:HPR})) leads to the formula 
\be \label{eq:cl}
\sin^2 \theta_{12} \simeq \frac 12 - \cos \phi \, \cos \theta_{23} \, 
\lambda_{12} ~,
\ee 
where $\lambda_{12}$ is the leading 12-rotation in $U_\ell$. 
From the experimentally observed $\sin^2 \theta_{12} \simeq 0.3$ 
it follows for the natural value of $\lambda_{12} \simeq 0.2$ that  
$\cos \phi$ has to be tuned to be very close to one. 
Leptonic CP violation 
in oscillation experiments is proportional to $\sin \phi$ 
\cite{lelmlt1,FPR} and very much suppressed. 

\end{itemize}

All these fine-tuning problems occurring in $L_e - L_\mu - L_\tau$ 
do not occur in scaling, which therefore represents a presumably 
better Ansatz for the inverted hierarchy.  Moreover, as we will 
elaborate upon in the next Section, scaling can easily be obtained 
in models based on discrete flavor symmetries, which are 
currently intensively studied \cite{dis_rev}.

\section{\label{sec:mod}A Model for Scaling and Phenomenological 
Consequences}

We consider now a seesaw model based on the 
$D_4 \times Z_2$ flavor symmetry which was proposed in Ref.~\cite{MR} 
to generate scaling. 
The particle content together with the 
quantum numbers under $D_4 \times Z_2$ is shown in Table \ref{tab:prop}.
The superscripts $+,-$ refer to the transformation under $Z_2$ and 
the rest are the $D_4$ representations. For the mathematical details 
of the $D_4$ group see for instance Ref.~\cite{CR}. 
Apart from the usual Majorana neutrinos $N_{e,\mu,\tau}$, the 
right-handed charged leptons $e_R, \mu_R, \tau_R$ and the lepton doublets 
$L_{e, \mu, \tau}$, one has to introduce five Higgs 
doublets $\phi_{1,2,3,4,5}$. In the Appendix we show as a proof of 
principle that the $D_4\times Z_2$-invariant Higgs potential 
can be minimized with Higgs masses having values above current limits. 
From the assignment 
in Table \ref{tab:prop} the following Lagrangian is obtained:  
\bea
{\cal L} = 
k_1 \, \overline{e_R} \, \phi_1 \, L_e + \overline{\mu_R} 
\left( 
k_2 \, \phi_1 - k_3 \, \phi_3 
\right) L_\mu + 
\overline{\tau_R} 
\left( 
k_2 \, \phi_1 + k_3 \, \phi_3 
\right) L_\tau \\[0.2cm]
+ h_1 \, \overline{N_e} \, \phi_1 \, L_e + h_2 \, \overline{N_\mu} 
\, \phi_2 \, L_e + h_3 \, \overline{N_\mu} 
\left( 
\phi_4 \, L_\tau + \phi_5 \, L_\mu 
\right) \\[0.2cm]
+ \frac 12 \left(
\overline{N_e} \, N_e^c \, M_1 + \overline{N_\mu} \, N_\mu^c\, M_2 + 
\overline{N_\tau} \, N_\tau^c \, M_3  
\right) + h.c.
\eea
Hence, the neutrino Dirac mass 
matrix can be written as 
\be \label{eq:md}
M_D = 
\left( 
\bad
a \, e^{i \varphi} & 0 & 0 \\[0.2cm]
b & d & e \\[0.2cm]
0 & 0 & 0 
\ea
\right) \, v_{\rm wk}~, 
\ee
and both the charged lepton and Majorana mass matrix 
are diagonal\footnote{There is another model based 
on $D_4\times Z_2$ presented in Ref.~\cite{MR}. The heavy Majorana mass 
matrix $M_R$ is arbitrary in this case, which has therefore 
little predictivity for leptogenesis.}. 
\begin{table}[ht]
\begin{center} 
\begin{tabular}{|c||c|} \hline 
Field & $D_4\times Z_2$ quantum number\\ \hline \hline
$L_e$ & $1_1^+$\\ \hline 
$e_R$, $N_e$, $ \phi_1$ & $1_1^-$ \\ \hline 
$N_\mu$, $\phi_2$ & $1_2^+$ \\ \hline 
$N_\tau$ & $1_2^-$ \\ \hline 
$\phi_3$ & $1_4^-$ \\ \hline
$\left(\begin{array}{c}L_\mu\\ L_ \tau\end{array}\right)$, 
$\left(\begin{array}{c}\phi_4 \\  \phi_5 \end{array}\right)$ 
 & $2^+$\\ \hline
$\left(\begin{array}{c}\mu_R\\  \tau_R\end{array}\right)$ & $2^-$\\ 
\hline 
\end{tabular}
\caption{\label{tab:prop}Transformation properties under $D_4 \times Z_2$ 
of the particle content of the model.}
\end{center} 
\end{table}
Note that multi-Higgs models as the one 
analyzed here typically predict flavor changing neutral currents 
and LFV in the charged lepton sector at 
dangerous levels. Here the model has a 
diagonal charged lepton mass matrix which renders 
processes like $\mu \ra e \gamma$ suppressed either 
by the GIM mechanism or by the masses of the heavy right-handed neutrinos. 
Note that the model presented here is non-supersymmetric. 
A supersymmetric version would have LFV via 
off-diagonal slepton mass matrices generated by the mechanism 
described in Section \ref{sec:gen} (see Eq.~(\ref{eq:BRs})). Because 
$M_D$ from Eq.~(\ref{eq:md}) is a special case (recall that 
$e = d/c$) of the form given in Eq.~(\ref{eq:md0}) one would in 
this case obtain the characteristic relation 
between the LFV charged lepton decays and atmospheric neutrino mixing 
from Eq.~(\ref{eq:LFV_rel}).

In $M_D$ shown in Eq.~(\ref{eq:md}) we have already included one complex 
phase. It is easy to show that with diagonal charged lepton and Majorana 
mass matrices there is only one complex phase in the model. 
Using all this, we can 
calculate the neutrino mass matrix using the type I seesaw 
formula to obtain 
\begin{eqnarray} \label{eq:mnumod}
{\cal M}_\nu~=~-\frac{v_{\rm wk}^2}{M_2} 
\left(
\begin{array}{ccc} \D 
\frac{M_2}{M_1} \, a^2 \, e^{2 i \varphi} + b^2 
& b\, d & b\, e \\[0.2cm]
b\, d & d^2  & d \, e \\[0.2cm]
b\, e & d \, e   &  e^2
\end{array} 
\right)~.
\end{eqnarray}
Note that the third heavy neutrino mass $M_3$ does not 
appear in ${\cal M}_\nu$, i.e., effectively we are dealing with 
a $2\times3$ seesaw. 
The low energy mass matrix apparently obeys scaling with $c = d/e$. 
We therefore have 
$m_3 = U_{e3}=0$ and $\tan^2 \theta_{23} = e^2 /d^2$. 
By evaluating the rephasing invariant ${\rm Im} \left\{ m_{ee} 
\, m_{\tau\tau} \, 
m_{e \tau}^\ast \, 
m_{\tau e}^\ast \right\}$ \cite{utpal} both with Eq.~(\ref{eq:mnumod}) 
and with the usual parametrization of the PMNS matrix,  
one finds a compact 
relation between the parameters in $M_D$ and $M_R$ and 
the low energy observables:  
\be \label{eq:compact}
\frac 14 \, \dma \, \dms \, 
\sin^4 \theta_{23} \, \sin^2 2 \theta_{12} \, \sin 2\alpha = 
\left(\frac{v_{\rm wk}^2}{M_2}\right)^4 \, 
\frac{M_2}{M_1} \, a^2 \, b^2 \, e^4 
\, \sin 2 \varphi ~.
\ee
The effective mass governing neutrinoless double beta decay is 
\be
\meff = \frac{v_{\rm wk}^2}{M_2} \, \left 
|\frac{M_2}{M_1} \, a^2 \, e^{2 i \varphi} + b^2 \right| = 
\sqrt{\dma} \, \sqrt{1 - \sin^2 2 \theta_{12} \, 
\sin^2 \alpha }~.
\ee

Let us focus now on high energy phenomenology in terms of 
leptogenesis \cite{lepto}. The decay asymmetries of 
the heavy neutrinos $N_{1,2,3}$ into final states with flavor 
$\alpha = e, \mu, \tau$ are \cite{flavor_flav}
\bea 
\varepsilon_i^\alpha \D  
= \frac{\D \Gamma (N_i \ra \phi \, \bar{l}_\alpha) -
\Gamma (N_i \ra \phi^\dagger \, l_\alpha)}
{\D \Gamma (N_i \ra \phi \, \bar{l}_\alpha) + 
\Gamma (N_i \ra \phi^\dagger \, l_\alpha)}  \\
= \D \frac{1}{8 \pi \, v_{\rm wk}^2} \, 
\frac{1}{(M_D \, M_D^\dagger)_{ii}}  
\, \sum\limits_{j \neq i} 
{\rm Im} \left\{ (M_D)_{i \alpha} \, 
(M_D^\dagger)_{\alpha j} \, 
\left(M_D \, M_D^\dagger 
\right)_{i j} \right\} \, f(M_j^2/M_i^2) ~, \\ 
\mbox{ where }   \D f(x) = 
\sqrt{x} \, \left(1 + 
\frac{1}{1 - x} - (1 + x) \, \ln \left( \frac{1+x}{x} \right) 
 \right) ~.
\eea 
With the very restricted form of the Dirac mass matrix from 
Eq.~(\ref{eq:md}) it turns out that only two of the nine possible 
$\varepsilon_i^\alpha$ are non-zero. 
Those are 
\bea \label{eq:eps} 
\D \varepsilon_1^e = \frac{1}{8 \pi} \, 
b^2 \, \sin 2 \varphi ~ f(M_2^2/M_1^2) 
\simeq -\frac{3}{16 \pi} \, b^2 \,\sin 2 \varphi~\frac{M_1}{M_2}
~,\\[0.2cm]
\D \varepsilon_2^e = \frac{-1}{8 \pi } \, 
\frac{a^2 \, b^2 }{b^2 + d^2 + e^2} \, \sin 2 \varphi ~
f(M_1^2/M_2^2) \simeq \frac{3}{16 \pi} \, 
\frac{a^2 \, b^2 }{b^2 + d^2 + e^2} \, \sin 2 \varphi 
~\frac{M_1}{M_2}~,
\eea
where we also gave the limits for $M_2^2 \gg M_1^2$. 
Again the third heavy neutrino does not play any role. 
The leptogenesis phase $\varphi$ is identical to the low energy 
Majorana phase $\alpha$, so that we can expect a correlation between 
the baryon asymmetry and the effective mass governing 
neutrinoless double beta decay. 
Let us focus in the following on the very typical case that 
$\varepsilon_1 = \varepsilon_1^e $ governs the baryon asymmetry. 
We need to specify the magnitude of the heavy neutrino masses. 
The common mass scale $v_{\rm wk}^2/M_2$ in the light neutrino mass matrix 
from Eq.~(\ref{eq:mnumod}) should be $\sqrt{\dma} \simeq 0.05$ eV, 
which brings us to the choice 
$M_2 = (5 \cdot 10^{13} \div 5 \cdot 10^{15})$ GeV, 
when the entries of the mass matrix lie between 0.1 and 10. 
We further note that in case of an inverted hierarchy all entries of 
${\cal M}_\nu$ are of the same order of magnitude, so the 
ratio $M_2/M_1$ should not be too large. Nevertheless, there should 
be a moderate hierarchy in order to avoid the complications and 
tuning issues of heavy neutrinos close in mass, so for definiteness we 
choose $M_1 = (\frac 16 \div \frac 13) \, M_2$. 
\begin{figure}[ht]
\begin{center}
\epsfig{file=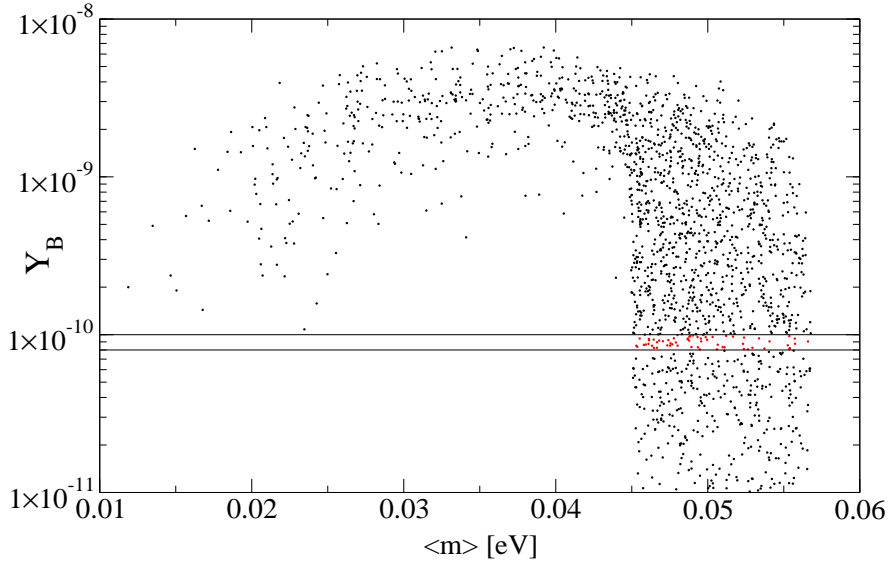,width=9cm,height=13cm,angle=270}
\caption{\label{fig:phen}Scatter plot of the effective mass against the 
baryon asymmetry of the Universe in the model from Section \ref{sec:mod}.}
\end{center}
\end{figure}
This in turn means that flavor effects 
in leptogenesis \cite{flavor_flav} do not play a role. In this limit, 
we can estimate the baryon asymmetry as 
\be
Y_B \simeq c_{\rm SP} \, \frac{\varepsilon_1}{g^\ast} \, \kappa(\tilde m_1)
~,
\ee
where $g^\ast = 122.75$, $c_{\rm SP} = -44/87$ 
and $\kappa$ can be parameterized as \cite{kappa} 
\[ 
\kappa(\tilde m_1) =  \left( 
\frac{0.55 \cdot 10^{-3}~\rm eV}{\tilde m_1}
\right)^{1.16}~,
\] 
with $\tilde m_1 = (M_D \, M_D^\dagger)_{11}/M_1$. In Fig.~\ref{fig:phen} 
we show the results of an analysis in which we searched with 
the ranges of $M_1$ and $M_2$ specified above for 
values of $a,b,d,e$ and $\phi$ which generate the correct neutrino 
mixing phenomenology as defined in Eq.~(\ref{eq:data}). Having 
found such parameters we evaluate the baryon asymmetry, 
which should lie in the range 
$(8 \div 10) \cdot 10^{-11}$ \cite{wmap}. 
We see that there is as expected a correlation between 
$Y_B$ and \meff~and that several of the points generate the correct 
baryon asymmetry, both in sign and magnitude. The particular choice 
of the parameters demands for successful leptogenesis 
the effective mass to lie around its maximal 
allowed value, $\meff \gs 0.045$ eV.

\section{\label{sec:concl}Summary}
In summary, we present a detailed investigation of the hypothesis that 
 the Majorana neutrino mass matrix obeys a scaling law as a way
to understand current neutrino observations. Two consequences of this 
hypothesis are that (i) the neutrino mass ordering is inverted 
rather than 
normal and (ii) both $U_{e3}$ and the lightest neutrino mass vanish. 
These results are 
invariant under renormalization group extrapolation and are therefore 
stable under radiative corrections, which distinguishes the scaling 
proposal from many others motivated by family symmetries or texture 
zeros. The effective mass governing neutrinoless double beta decay 
can be as large as $\sqrt{\dma} \simeq 0.06$ eV. 
Another distinguishing prediction of scaling is that 
the value of the atmospheric mixing angle is not necessarily maximal 
even though $U_{e3}=0$. This is in contrast to 
models with $\mu$--$\tau$ symmetry, which provide 
a simple way to understand both maximal atmospheric mixing with 
a very small $U_{e3}$. Models with approximate or broken 
$\mu$--$\tau$ symmetry always correlate 
non-vanishing of $U_{e3}$ with deviations from maximal $\theta_{23}$. 
We note that recent analyzes of the available neutrino data tend to favor 
non-maximal atmospheric neutrino mixing. 
We also compare the 
scaling hypothesis to the $L_e - L_\mu - L_\tau$ flavor symmetry which is 
very often utilized to understand an inverted hierarchy. 
Unlike the scaling Ansatz, fitting observations 
requires a fine-tuning of mass matrix elements/perturbation parameters. 
An interesting aspect of our hypothesis that it is invariant under any 
possible rotation of the basis in $\mu$--$\tau$ space, e.g., 
coming from the charged lepton sector. This may make it easier to 
construct models that obey scaling. We also note 
ways to test scaling using high energy 
astrophysical neutrino fluxes. We discuss a particular 
class of seesaw models based on the 
$D_4 \times Z_2$ flavor symmetry that realize the scaling hypothesis. 
There are (i) no dangerous flavor changing neutral currents in the 
lepton sector, (ii) the leptogenesis phase is identical to the 
low energy Majorana phase and (iii) successful leptogenesis requires the  
effective mass in neutrinoless double beta decay to be larger than 45 meV.

\vspace{0.3cm}
\begin{center}
{\bf Acknowledgments}
\end{center}
R.N.M.~was supported by the National Science Foundation 
grant no.~Phy--0354401 and by the 
Alexander von Humboldt Foundation (the Humboldt Research Award). 
The work was also supported by the 
EU program ILIAS N6 ENTApP WP1 and by the 
Deutsche Forschungsgemeinschaft 
in the DFG-Sonderforschungsbereich 
Transregio 27 ``Neutrinos and beyond -- 
Weakly interacting particles in 
Physics, Astrophysics and Cosmology'' (A.B.~and W.R.), 
and under project number RO--2516/3--2 (W.R.). A.B.~acknowledges 
support from the Studienstiftung des Deutschen Volkes.

\renewcommand{\theequation}{A\arabic{equation}}
\setcounter{equation}{0}

\begin{appendix}
\section{Scalar Potential}
\label{sec:potential}

The model we study in Section \ref{sec:mod} is defined by 
the transformation 
properties given in Table \ref{tab:prop}. It includes five 
Higgs doublets $\phi_1$, $\phi_2$, $\phi_3$, $\phi_4$ and $\phi_5$. 
We will show now that the potential involving all these fields  
can be minimized with a realistic vev configuration leading to a 
phenomenologically viable spectrum of scalar bosons below mass bounds 
from direct production. 
Five Higgs fields correspond to ten charged and ten neutral ones, or to  
20 physical degrees of freedom. 
Of the ten charged degrees of freedom, two are eaten by the 
electroweak $W$ bosons, leaving eight degrees of freedom, i.e., 
four complex charged scalars. Of the ten neutral degrees of 
freedom five are even under charge conjugation. These give five 
real physical scalars. The other five neutral degrees of 
freedom are odd under charge conjugation. One of these is eaten 
by the electroweak $Z$ boson, leaving four physical pseudoscalars. 
With the assignment given in Table \ref{tab:prop} the 
most general $D_4\times Z_2$-invariant potential with real 
coefficients reads: 
\begin{eqnarray}
\label{eq:potential}
V & = &- \sum_{i=1}^3 \mu_i^2 \phi_i^\dagger \phi_i - 
\mu_4^2 (\phi_4^\dagger \phi_4 + \phi_5^\dagger \phi_5) 
+ \sum_{i=1}^3 \lambda_i (\phi_i^\dagger \phi_i)^2 + 
\lambda_4 (\phi_4^\dagger \phi_4 + \phi_5^\dagger \phi_5)^2 \nonumber\\
&+& \lambda_{12} (\phi_1^\dagger \phi_1)(\phi_2^\dagger \phi_2) 
+ \lambda_{13} (\phi_1^\dagger \phi_1)(\phi_3^\dagger \phi_3) 
+ \lambda_{23} (\phi_2^\dagger \phi_2)(\phi_3^\dagger \phi_3) \nonumber\\
&+& \sum_{i=1}^3 \kappa_i (\phi_i^\dagger \phi_i) 
(\phi_4^\dagger \phi_4 + \phi_5^\dagger \phi_5) 
+  \alpha_1 [(\phi_1^\dagger \phi_2)^2 + h.c.] 
+ \alpha_2 \vert \phi_1^\dagger \phi_2 \vert^2 \\ 
&+& \alpha_3 [(\phi_2^\dagger \phi_3)^2 + h.c.] 
+ \alpha_4 \vert \phi_2^\dagger \phi_3 \vert^2 
+ \alpha_5 (\phi_4^\dagger \phi_5 + \phi_5^\dagger \phi_4)^2 
+ \alpha_6 (\phi_4^\dagger \phi_5 - \phi_5^\dagger \phi_4)^2 \nonumber\\
&+& \alpha_7 (\phi_4^\dagger \phi_4 - \phi_5^\dagger \phi_5)^2 
+ \alpha_8 (\phi_4^\dagger \phi_4 - \phi_5^\dagger \phi_5) 
(\phi_1^\dagger \phi_3 + h.c.) + \alpha_9 [(\phi_2^\dagger \phi_4)^2 
+ (\phi_2^\dagger \phi_5)^2 + h.c.] \nonumber\\
&+& \alpha_{10} (\vert \phi_2^\dagger \phi_4 \vert^2 
+ \vert \phi_2^\dagger \phi_5 \vert^2) \nonumber\\
&+& \alpha_{11} [(\phi_1^\dagger \phi_4)^2 
+ (\phi_1^\dagger \phi_5)^2 + h.c.] + \alpha_{12} 
(\vert \phi_1^\dagger \phi_4 \vert^2 + 
\vert \phi_1^\dagger \phi_5 \vert^2) \nonumber\\
&+& \alpha_{13} [(\phi_3^\dagger \phi_4)^2 
+ (\phi_3^\dagger \phi_5)^2 + h.c.] + 
\alpha_{14} ( \vert \phi_3^\dagger \phi_4 \vert^2 
+ \vert \phi_3^\dagger \phi_5 \vert^2 ) \nonumber\\
&+& \alpha_{15} [ (\phi_1^\dagger \phi_4)(\phi_3^\dagger \phi_4) 
- (\phi_1^\dagger \phi_5)(\phi_3^\dagger \phi_5) + h.c.] 
+ \alpha_{16} [(\phi_1^\dagger \phi_4)(\phi_4^\dagger \phi_3) 
- (\phi_1^\dagger \phi_5)(\phi_5^\dagger \phi_3) + h.c.]\nonumber
\end{eqnarray}
This potential can for instance be minimized by choosing 
the VEV configuration:
\begin{equation}
\label{eq:positiveVEV}
\langle \phi_i \rangle = \left( \begin{array}{c} 0 
\\ \frac{v_{\rm wk}}{\sqrt{5}} \end{array} \right)~\mbox{ for } 
i=1,2,4,5~\mbox{ and } 
\langle \phi_3 \rangle = \left( \begin{array}{c} 0 \\ 
- \frac{v_{\rm wk}}{\sqrt{5}} \end{array} \right)~,
\end{equation}
if one uses the following numerical values for the coefficients:
\begin{eqnarray}
\label{eq:coefficients}
\lambda_1 = 2.62904\, ; &  \lambda_2 = 2.91805\, ; &  
\lambda_3 = 2.53936 \, ;\nonumber\\
\lambda_4 = 2.93754\, ; &  \lambda_{12} = 2.94576\, ; &  
\lambda_{13} = -1.29472 \, ;\nonumber\\
\lambda_{23} = -0.455254\, ; &  \kappa_1 = 0.190918\, ; &  
\kappa_2 = 2.25757 \, ;\nonumber\\
\kappa_3 = -0.778689\, ; &  \alpha_1 = -1.58251\, ; &  
\alpha_2 = -0.77218 \, ;\nonumber\\
\alpha_3 = -2.649\, ; &  \alpha_4 = 2.32869\, ; &  
\alpha_5 = 0.690183 \, ;\\ %\nonumber\\
\alpha_6 = -1.82076\, ; &  \alpha_7 = -0.097017\, ; &  
\alpha_8 = -1.38637 \, ;\nonumber\\
\alpha_9 = -0.853325\, ; &  \alpha_{10} = 0.393294\, ; &  
\alpha_{11} = -0.404394 \, ;\nonumber\\
\alpha_{12} = 0.68049\, ; &  \alpha_{13} = -1.29704\, ; &  
\alpha_{14} = 0.764254 \, ;\nonumber\\
\alpha_{15} = 1.70881\, ; &  \alpha_{16} = -0.322439\, ; &  
\mu_1^2 = 0.619434 \, v_{\rm wk}^2 \, ;\nonumber\\
\mu_2^2 = 0.661705 \, v_{\rm wk}^2\, ; &  \mu_3^2 = -0.971516 
\, v_{\rm wk}^2\, ; &  \mu_4^2 = 1.4775 \, v_{\rm wk}^2  \ .\nonumber 
\end{eqnarray}
One finds masses of 552 GeV, 390 GeV, 362 GeV, 240 GeV and 147 GeV 
for the scalars, 682 GeV, 607 GeV, 460 GeV and 
322 GeV for the pseudoscalars and 404 GeV, 363 GeV, 273 GeV 
and 155 GeV for the charged scalars. These scalars are in general 
superpositions of all five Higgs fields, except for the 240 GeV scalar, 
which is only a linear combination of $\phi_4$ and $\phi_5$. 
Note that our model is only focussing on the lepton sector. 
Confronting the obtained Higgs mass values with limits stemming from rare 
meson decays would mean to construct a full 
model including also the quark sector and carefully perform a lengthy 
study of the diagrams leading to flavor changing neutral currents taking 
into account all five Higgs doublets. 
Within our model the specific choice of 
Eq.~(\ref{eq:positiveVEV}) leads 
in the Dirac mass matrix from Eq.~(\ref{eq:md}) to $d = e$ and therefore 
maximal atmospheric neutrino mixing. Other values are of course 
possible, which would lead to slightly different scalar masses and 
parameters in Eq.~(\ref{eq:coefficients}). 

\end{appendix}

\end{document}